\documentclass{article}
\usepackage{graphicx}
\usepackage{url}      
\usepackage{hyperref} 

\usepackage[authoryear]{natbib}

\bibpunct{[}{]}{;}{a}{ }{,}

\title{Cancer Survival Rates Are Misleading}
\author{Allen Downey\\
Olin College of Engineering\\
1000 Olin Way\\
Needham, MA 02492\\
\href{mailto:allen.downey@olin.edu}{allen.downey@olin.edu}}
\date{\today}

\begin{document}

\maketitle

\section{Introduction}

Five-year survival rates might be the most misleading statistics in medicine. 
For example, suppose 5-year survival for a hypothetical cancer is

\begin{itemize}
    \item 91\% among patients diagnosed early, while the tumor is localized at the primary site,
    \item 74\% among patients diagnosed later, when the tumor has spread regionally to nearby lymph nodes or adjacent organs, and
    \item 16\% among patients diagnosed late, when the tumor has spread to distant organs or lymph nodes.
\end{itemize}

What can we infer from these statistics?

\begin{itemize}
    \item If a patient is diagnosed early, it is tempting to think the probability is 91\% that they will survive five years after diagnosis.
    
    \item Looking at the difference in survival between early and late detection, it is tempting to conclude that more screening would save lives.
    
    \item In a case where a patient is diagnosed late and dies of cancer, it is tempting to say that they would have survived if their cancer had been caught early.
    
    \item And if 5-year survival increases over time, it is tempting to conclude that treatment has improved.
\end{itemize}

In fact, none of these inferences are correct.
To see why, we'll use a simple Markov model of tumor progression, which shows that the patterns we see in real survival rates---higher survival rates with early detection, and improvement over time---can appear even if early detection has no benefit and treatment is completely ineffective.
Of course, that's an extreme example, but it shows why survival rates alone do not support these inferences.

Let's take them one at a time.

\section{Particularization}

Here's the first incorrect inference:

\begin{quote}
If a patient is diagnosed early, and 5-year survival after early diagnosis is 91\%, it is tempting to think the probability is 91\% that they will survive five years after diagnosis.
\end{quote}

This is almost correct in the sense that it applies to the past cases that were used to estimate the survival rate---of all patients in the dataset who were diagnosed early, 91\% of them survived at least five years.

But it is misleading for two reasons:

\begin{itemize}
    \item Because it is based on past cases, it doesn't apply to present cases if the effectiveness of treatment has changed or---often more importantly---if diagnostic practices have changed.

    \item Also, before interpreting a probability like this, which applies in general, it is important to particularize it for a specific case.
\end{itemize}

Factors that should be taken into account include the general health of the patient, their age, and the mode of detection. Some factors are causal---for example, general health directly improves the chance of survival. Other factors are less obvious because they are informational---for example, the mode of detection can make a big difference:

\begin{itemize}
    \item If a tumor is discovered because it is causing symptoms, it is more likely to be larger, more aggressive, and relatively late for a given stage---and all of those implications decrease the chance of survival.

    \item If a tumor is not symptomatic, but discovered during a physical exam, it is probably larger, later, and more likely to cause mortality, compared to one discovered by high resolution imaging or a sensitive chemical test.

    \item Conversely, tumors detected by screening are more likely to be slow-growing because of length-biased sampling---the probability of detection depends on the time between when a tumor is detectable and when it causes symptoms.
\end{itemize}

Taking age into account is complicated because it might be both causal and informational, with opposite implications. A young patient might be more robust and able to tolerate treatment, but a tumor detectable in a younger person is likely to have progressed more quickly than one that could only be discovered after more years of life. So the implication of age might be negative among the youngest and oldest patients, and positive in the middle-aged.

For some cancers, the magnitude of these implications is large, so the probability of 5-year survival for a particular patient might be higher than 91\% or much lower.

\section{Is More Screening Better?}

Now let's consider the second incorrect inference.

\begin{quote}
If 5-year survival is high when a cancer is detected early and much lower when it is detected late, it is tempting to conclude that more screening would save lives.
\end{quote}

For example, in a recent video \citep{gal2025video}, Nassim Taleb and Emi Gal discuss the pros and cons of cancer screening, especially full-body MRIs for people who have no symptoms. They note that survival is highest if a tumor is detected while localized at the primary site, lower if it has spread regionally, and often much lower if it has spread distantly.

They take this as evidence that screening for these cancers is beneficial. For example, at one point Taleb says, ``Look at the payoff for pancreatic cancer---10 times the survival rate.''

And Gal adds, ``Colon cancer, it's like seven times\ldots{} The overarching insight is that you want to find cancer early\ldots{} This table makes the case for the importance of finding cancer early.''

Taleb agrees, but this inference is incorrect: these statistics do not make the case that it is better to catch cancer early.

Catching cancer early is beneficial only if (1) the cancers we catch would otherwise cause disease and death, (2) we have treatments that prevent those outcomes, and (3) the benefits outweigh the costs of additional screening. Survival rates do not show that any of those things is true.

In fact, it is possible for a cancer to reproduce the survival rates we observe, even if we have no treatment and detection has no effect on outcomes. To demonstrate, I'll use a model of tumor progression to show that a hypothetical cancer could have the same survival rates as colon cancer---even if there is no effective treatment.

To be clear, I'm not saying that cancer treatment is not effective---in many cases we know that it is. I'm saying that we can't tell, just looking at survival rates, whether early detection has any benefit at all.

We'll use data from the Surveillance, Epidemiology, and End Results (SEER) Program \citep{seerdata}, which collects cancer statistics from population-based registries covering about 28\% of the U.S. population. The following table shows 5-year survival rates for the most common cancer sites, based on diagnoses from 2014-2020.

\begin{table}[h]
\centering
\caption{5-year survival rates by stage at diagnosis (2014-2020)}
\begin{tabular}{lccc}
\hline
Cancer Site & Localized & Regional & Distant \\
\hline
Breast & 99.6\% & 86.8\% & 32.2\% \\
Lung and Bronchus & 63.6\% & 36.5\% & 8.7\% \\
Colon and Rectum & 91.4\% & 74.0\% & 15.8\% \\
Melanoma of the Skin & 100.0\% & 73.9\% & 33.2\% \\
Non-Hodgkin Lymphoma & 86.2\% & 78.1\% & 67.7\% \\
Kidney and Renal Pelvis & 93.4\% & 75.1\% & 17.7\% \\
\hline
\end{tabular}
\label{tab:survival}
\end{table}

In general, survival rates are highest when cancers are detected at an early (localized) stage, and lowest when detected at a late (distant) stage. But that doesn't necessarily mean that early detection is beneficial.

\subsection{Markov Chain}

To see why not, we'll model tumor progression using a discrete-time Markov chain with these states:

\begin{itemize}
    \item U1, U2, and U3 represent tumors that are undetected at each stage: local, regional, and distant.
    \item D1, D2, D3 represent tumors that were detected/diagnosed at each stage.
    \item And M represents mortality.
\end{itemize}

The following figure shows the states and transition rates.

\begin{figure}[h]
    \centering
    \includegraphics[width=0.85\textwidth]{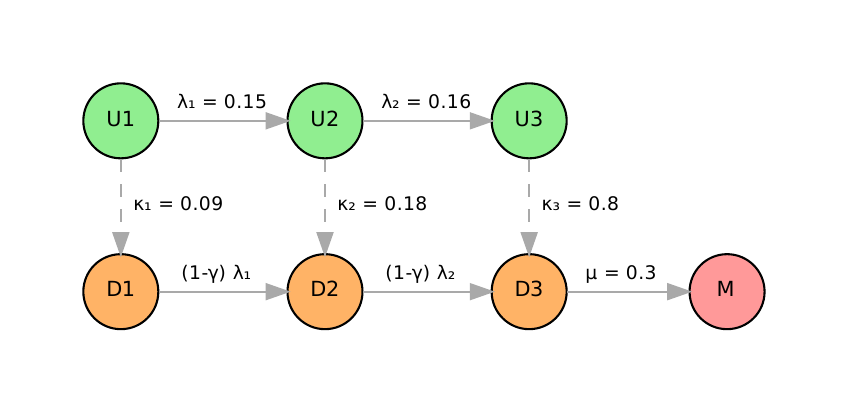}
    \caption{States and transition rates of the model}
    \label{fig:model}
\end{figure}

The transition probabilities are:

\begin{itemize}
    \item $\lambda_i$, two values that represent transition rates between stages,
    \item $\kappa_j$, three values that represent detection rates at each state,
    \item $\mu$, the mortality rate from D3,
    \item $\gamma$, the effectiveness of treatment.
\end{itemize}

If $\gamma > 0$, the treatment is effective by decreasing the probability of progression to the next stage. In the models we'll run for this example, $\gamma=0$, which means that detection has no effect on progression---it only starts the clock for 5-year survival.

Overall, these values are within the range we observe in real cancers.

\begin{itemize}
    \item The transition rates are close to $1/6$, which means the average time at each stage is $6$ simulated years.
    \item The detection rate is low at the first stage, higher at the second, and much higher at the third.
    \item The mortality rate is close to $1/3$, so the average survival after diagnosis at the third stage is about $3$ years.
\end{itemize}

In the model, death occurs only after a cancer has progressed to the third stage and been detected. That's not realistic---in reality deaths can occur at any stage, due to cancer or other causes.

But adding more transitions would not make the model better. The purpose of the model is to show that we can reproduce the survival rates we see in reality, even if there are no effective treatments. Making the model more realistic would increase the number of parameters, which would make it easier to reproduce the data, but that would not make the conclusion stronger.

\subsection{Simulation}

Now we can use the model to simulate tumor progression.
From the rates in the previous diagram we can compute Table~\ref{tab:transition}, which shows the probability of moving from each state (row) to each other state (column) in one simulated year.

\begin{table}[h]
\centering
\caption{Transition matrix (one-year probabilities)}
\label{tab:transition}
\begin{tabular}{lccccccc}
\hline
 & U1 & U2 & U3 & D1 & D2 & D3 & M \\
\hline
U1 & 0.76 & 0.15 & 0.00 & 0.09 & 0.00 & 0.00 & 0.00 \\
U2 & 0.00 & 0.66 & 0.16 & 0.00 & 0.18 & 0.00 & 0.00 \\
U3 & 0.00 & 0.00 & 0.20 & 0.00 & 0.00 & 0.80 & 0.00 \\
D1 & 0.00 & 0.00 & 0.00 & 0.85 & 0.15 & 0.00 & 0.00 \\
D2 & 0.00 & 0.00 & 0.00 & 0.00 & 0.84 & 0.16 & 0.00 \\
D3 & 0.00 & 0.00 & 0.00 & 0.00 & 0.00 & 0.70 & 0.30 \\
M  & 0.00 & 0.00 & 0.00 & 0.00 & 0.00 & 0.00 & 1.00 \\
\hline
\end{tabular}
\end{table}

For example, if a simulated person is in state U1, the probability is 0.15 that they transition to U2, 0.09 that they transition to D1, and 0.76 that they stay in U1.
Notice that each row sums to 1.

We simulate the progression of 10,000 tumors, each starting in U1. 
At each time step, we use the probabilities for the current state to choose the next state at random.
We run 100 steps, long enough that all simulations reach state M, and record the trajectory of consecutive states.
For each trajectory, we record the stage at time of diagnosis and whether death occurs within five years after diagnosis.
Then we compute the percentage of cases diagnosed at each stage and 5-year survival rates conditioned on stage.

Here are the results:

\begin{itemize}
    \item In the model, survival rates are 95\% if localized, 72\% if spread regionally, and 17\% if spread distantly. For colon cancer, the rates from SEER data are 91\%, 74\%, and 16\%. So the simulation results are not exactly the same, but they are close.
    
    \item In the model, 38\% of tumors are localized when diagnosed, 33\% have spread regionally, and 29\% have spread distantly. The actual distribution for colon cancer is 38\%, 38\%, and 24\%. Again, the simulation results are not exactly the same, but close.
\end{itemize}

With more trial and error, we could probably find parameters that reproduce the results exactly. After all, the model has seven parameters and we are matching observations with only five degrees of freedom.

That might seem unfair, but it makes the point that there is not enough information in the survival table---even if we also consider the distribution of stages---to estimate the parameters of the model. The data don't exclude the possibility that treatment is ineffective, so they don't prove that early detection is beneficial.

Again, it \textit{might} be better to find cancer early, if the benefit of treatment outweighs the costs of false discovery and overdiagnosis---but that's a different analysis, and 5-year survival rates aren't part of it.

\section{Counterfactuals}

Now let's consider the third incorrect inference.

\begin{quote}
In a case where a patient is diagnosed late and dies of cancer, it is tempting to say that they would have survived if their cancer had been caught early.
\end{quote}

For example, later in the previous video \citep{gal2025video}, Taleb says, ``Your mother, had she had a colonoscopy, she would be alive today\ldots{} she's no longer with us because it was detected when it was stage IV, right?'' And Gal agrees.

That might be true, if treatment would have prevented the cancer from progressing. But this conclusion is not supported by the data in the survival table.

If someone is diagnosed late and dies, it is tempting to look at the survival table and think there's a 91\% chance they would have survived if they had been diagnosed earlier. But that's not accurate---and it might not even be close.

First, remember what 91\% survival means: among people diagnosed early, 91\% survived five years after diagnosis. But among those survivors, an unknown proportion had tumors that would not have been fatal, even without treatment. Some might be non-progressive, or progress so slowly that they never cause disease or death. But in a case where the patient dies of cancer, we know their tumor was not one of those.

As a simplified example, suppose that of all tumors that are caught early, 50\% would cause death within five years, if untreated, and 50\% would not. Now imagine 100 people, all detected early and all treated: 50 would survive with or without treatment; out of the other 50, 41 survive with treatment---so overall survival is 91\%. But if we know someone is in the second group, their chance of survival is 41/50, which is 82\%.

And if the percentage of non-progressive cancers is higher than 50\%, the survival rate for progressive cancers is even lower, holding overall survival constant. So that's one reason the inference is incorrect.

To see the other reason, let's be precise about the counterfactual scenario. Suppose someone was diagnosed in 2020 with a tumor that had spread distantly, and they died in 2022. Would they be alive in 2025 if they had been diagnosed earlier?

That depends on when the hypothetical diagnosis happens---if we imagine they were diagnosed in 2020, five year survival might apply (except for the previous point). But if it had spread distantly in 2020, we have to go farther back in time to catch it early. For example, if it took 10 years to progress, catching it early means catching it in 2010. In that case, being ``alive today'' would depend on 15-year survival, not 5-year.

\subsection{It's a Different Question}

The five-year survival rate answers the question ``Of all people diagnosed early, how many survive five years?'' That is a straightforward statistic to compute.

But the hypothetical asks a different question: ``Of all people who died [during a particular interval] after being diagnosed late, how many would be alive [at some later point] if the tumor had been detected early?'' That is a much harder question to answer, and five-year survival provides little or no help.

In general, we don't know the probability that someone would be alive today, if they had been diagnosed earlier. Among other things, it depends on progression rates with and without treatment.

\begin{itemize}
    \item If many of the tumors caught early would not have progressed or caused death, even without treatment, the counterfactual probability would be low.
    
    \item In any case, if treatment is ineffective, as in the hypothetical cancer we simulated, the counterfactual probability is zero.
    
    \item At the other extreme, if treatment is perfectly effective, the probability is 100\%.
\end{itemize}

It might be frustrating that we can't be more specific about the probability of the counterfactual, but if someone you know was diagnosed late and died, and it bothers you to think they would have lived if they had been diagnosed earlier, it might be some comfort to realize that we don't know that---and it might be unlikely.

\section{Comparing Survival Rates}

Now let's consider the last incorrect inference.

\begin{quote}
If 5-year survival increases over time, it is tempting to conclude that treatment has improved.
\end{quote}

This conclusion is appealing because if cancer treatment improves, survival rates improve, other things being equal. But if we do more screening and catch more cancers early, survival rates improve even if treatment is no more effective. And if screening becomes more sensitive and detects smaller tumors, survival rates also improve.

For many cancers, all three factors have changed over time: improved treatment, more screening, and more sensitive screening. Looking only at survival rates, we can't tell how much change in survival we should attribute to each.

In 2000, Welch and colleagues noted \citep{welch2000increasing}, ``Although 5-year survival is a valid measure for comparing cancer therapies in a randomized trial, our analysis shows that changes in 5-year survival over time bear little relationship to changes in cancer mortality. Instead, they appear primarily related to changing patterns of diagnosis.''

That conclusion might be stated too strongly. A response paper \citep{lichtenberg2010increasing} concludes ``While the change in the 5-year survival rate is not a perfect measure of progress against cancer [\ldots{}] it does contain useful information; its critics may have been unduly harsh. Part of the long-run increase in 5-year cancer survival rates is due to improved [\ldots{}] therapy.''

But with survival rates alone, we can't say what part---and the answer is different for different sites. For example, this paper concludes \citep{cho2014changes}:

\begin{quote}
In some cases, increased survival was accompanied by decreased burden of disease, reflecting true progress. For example, from 1975 to 2010, five-year survival for colon cancer patients improved \ldots{} while cancer burden fell: Fewer cases \ldots{} and fewer deaths \ldots{}, a pattern explained by both increased early detection (with removal of cancer precursors) and more effective treatment. In other cases, however, increased survival did not reflect true progress. In melanoma, kidney, and thyroid cancer, five-year survival increased but incidence increased with no change in mortality. This pattern suggests overdiagnosis from increased early detection, an increase in cancer burden.
\end{quote}

A 2024 review explains \citep{welch2024overdiagnosis}: ``Overdiagnosis inflates both the denominator (the number diagnosed) and the numerator (the number alive 5 years later), causing 5-year survival to rise even if the number of deaths is unchanged.''

That's why I conclude, as Gigerenzer does \citep{gigerenzer2013five}: ``Only reduced mortality rates can prove that screening saves lives\ldots{} journal editors should no longer allow misleading statistics such as five year survival to be reported as evidence for screening.''

\section{Summary}

Survival rates are misleading because they suggest inferences they do not actually support.

\begin{itemize}
    \item Rates from the past might not apply to the present, and for a particular patient the probability of survival depends on (1) causal factors like general health, and (2) informational factors like the mode of discovery (screening vs symptomatic presentation).
    
    \item If survival rates are higher when tumors are discovered early, that doesn't mean that more screening would be better.

    \item And if a cancer is diagnosed late, and the patient dies, that doesn't mean that if it had been diagnosed early, they would have lived.

    \item Finally, if survival rates improve over time (or they are different in different places) that doesn't mean treatment is more effective.
\end{itemize}

To be clear, all of these conclusions \textit{can} be true---for some cancers, treatments have improved, and for some, additional screening would save lives. But to support these conclusions, we need other methods and metrics, notably randomized controlled trials that compare mortality. Survival rates alone provide little or no information, and they are more likely to mislead than inform.

For details of the Markov model and additional analysis, see the Jupyter notebook available at \url{https://allendowney.github.io/ThinkBayes2/cancer.html} or run it interactively on Colab at \url{https://colab.research.google.com/github/AllenDowney/ThinkBayes2/blob/master/examples/cancer.ipynb} \citep{downey2025notebook}.

\subsection{About the author}

Allen Downey is a professor emeritus at Olin College and a consultant specializing in data science and Bayesian statistics. He is the author of several books -- including Think Python, Think Bayes, and Probably Overthinking It -- and a blog about programming and data science. He received a Ph.D. in computer science from the University of California, Berkeley, and Bachelor's and Masters degrees from MIT. 

\bibliographystyle{plainnat}
\bibliography{main}

\end{document}